# Emerging Trends in Soybean Industry

*SIDDHARTHA PAUL TIWARI[1]*

*Google Asia Pacific Pte. Ltd., 70 Pasir Panjang Road,*
*Mapletree Business City II, Singapore 117371*
E mail: siddhartha@google.com

## ABSTRACT

*Soybean is the most globalized, traded and processed crop commodity. USA, Argentina and Brazil continue to be the top three producers and exporters of soybean and soymeal. Indian soy industry has also made a mark in the national and global arena. While soymeal, soyoil, lecithin and other soy-derivatives stand to be driven up by commerce, the soyfoods for human health and nutrition need to be further promoted. The changing habitat of commerce in soy derivatives necessitates a shift in strategy, technological tools and policy environment to make Indian soybean industry continue to thrive in the new industrial era. Terms of trade for soy farming and soy-industry could be further improved. Present trends, volatilities, slowdowns, challenges faced and associated desiderata are accordingly spelt out in the present article.*

## Introduction

Soybean is one of the leading crop commodities produced, traded and utilized globally. Barring some small portion, the bulk of it needs industrial processing and value addition. Besides its numerous uses, soybean is predominantly used as soymeal typically as source of high protein for animal feed and as edible oil. A close and essential association between soybean farming and soy-industry including food and feed industry makes soybean a perfect global trade commodity. However, its best benefits can be reaped by turning it into value-added products for human health and nutrition.

Soybean and industry cannot be sundered. As a glaring example, Henry Ford, the noted industrialist, had a deep and continued inclination for soybean as realized by his efforts and achievements in establishing soybean plants in 1930s, putting soybean products to various industrial uses and promoting soybean as a whole (Shurtleff and Aoyagi 2011). A world famous photo dated November 1940 shows Henry Ford, dressed in coat and hat, swinging an axe at the plastic trunk lid of a car made of soy fibre. Industrial world, through three main industrial revolutions, has undergone a paradigm shift from hand tools and basic machines to (i) powered mechanization,

(ii) mass production along assembly lines, and (iii) automation and globalization. The Fourth Industrial Revolution (4 IR) is unfolding in the present early 21st century and is characterized by IT-based modern technologies with supporting IT structures, robotization, artificial intelligence, cyber-physical systems and global value chains (Schwab, 2016/2017, Micklethwait and Wooldridge, 2014). Overarching impact of 4 IR is expected on physical, digital and biological world. The present article overviews the present trends of the soybean trade, industry and commerce in this new industrial era.

## Role of industry in Indian soy revolution

Saga of success of soy-revolution in India has been told earlier (Tiwari *et al.*, 1999) depicting the renaissance of traditional soybean to a commercial crop. For rapid spread of soybean, more than a score of causes and associated organisations have been identified (Tiwari *et al.*, 1999; Chand, 2007; Dupare *et al.*, 2008; Tiwari, 2014). This article does not intend a comprehensive coverage of soy-revolution and highlights the role of soy-industry alone. In early seventies, some businessmen of Indore in central India foresaw great profit in export of soymeal mainly to European countries. Soybean did not occupy a large area that time. Although a number of solvent extraction plants existed then, those were not used for getting soymeal, oil and lecithin out of soybean grains primarily because soybean's availability and its potential for oil and export of soymeal was not yet sizeably realized.

Entrepreneurship of Indian businessmen to go for investment, labour and risk is praiseworthy in starting the extraction and export of soymeal. A lot of credit goes to several businessmen of Indore particularly Shahras (Mahadeo Bhai and Kailash Shahra), N N Jain and several others. With the improved cultivation and availability of seeds, Shahras set up a 50 tonne per day (TPD) plant in 1973. This plant was modernized and upgraded in the next two years and, with the addition of more plants, the capacity increased. Soybean industry started gradually but soon took off with a boom. The farmers were paid fairly and that resulted in sharp increase in soybean area, filling-up rainy season fallow land in central India with Indore as its epicentre.

A little later in 1979, the Madhya Pradesh Oilseeds Growers' Federation (OILFED) was established. It had its own solvent extraction plants and was a foreign-exchange earner through export of soymeal. It was also one of the main organisations that helped spread the soy revolution by procuring soybean from farmers at a remunerative price to them and providing inputs and knowledge on an overall basis. National Dairy Development Board (NDDB), the Rajasthan Oilseed Growers' Federation (Tilam Sangh/RAJFED), the National Agricultural Cooperative Marketing Federation of India (NAFED) and the state marketing federations also promoted soybean in Madhya Pradesh and adjoining areas in Maharashtra and Rajasthan particularly through market intervention and price support. The Soybean Processors Association of India (SOPA) was also established in 1979 with its headquarters in Indore. Besides promoting soy-processing and export of meal and value-added soyfood products from India, it provides technical backstopping through its analytical and research laboratories. It also promotes larger usage of soy-foods for better health and nutrition. It undertook a Soybean Development Programme to help farmers particularly in adopting new technology and using improved quality seed.

Eventual commercial success of Indian soybean owes much to the concurrent development of the soy industry comprising the value chain from grain to solvent extraction plants to soymeal and other derivatives and export mechanism which provided remunerative market to soybean growers. The profit of the industry trickled down to farmers and other stakeholders. Soybean cultivation as a crop, thus, really began in India in the early seventies on an area of about 30,000 hectares. The acreage grew gradually and steadily over decades. The soybean crop presently covers an area of about 12 million hectares with a total production of about 14 million tonnes. The three largest soybean producing states are Madhya Pradesh, Maharashtra and Rajasthan (Directorate of Economics and Statistics, 2016). Owing to possessing and sustaining a major share of soybean area (about 57 % at present), Madhya Pradesh is called the ‗Soy-State'.

**Emulative Models supporting soybean industry as evinced by soybean revolution of India**

Soybean, despite its short stay as a crop in India (1970s and thereafter), has served as a base in establishing some distinctive models such as (1) futures exchange (Soybean Board of Trade (SBoT) now National Board of Trade Ltd., (NBoT), (2) global value chain and global trade, (3) use of ICTs (*e.g.* ITC's e Choupals/Soy-Choupals) towards the technology adoption and domestic trade facilitation, (4) farmer-friendly extension services by private sector (*e.g.* private extension of Dhanuka and participatory approach of ‗Gramin Vikas Trust'), and (5) development of Indore as a ‗neo-seed hub' dealing with high volume-low value seed. Some of these concomitant developments are briefly depicted below.

The NBoT is a commodity futures exchange especially for edible oil and oilseeds in India with needed automation, online trading platform in Indore, nation-wide screen-based trading system, *etc*. It was mainly due to soybean that Indore, the commercial capital of Madhya Pradesh, also developed as a neo-seed hub having several sizeable seed companies. The special feature of these seed companies is that they not only deal with low volume-high value crop hybrid and vegetable seed but also with a large amount of high volume-low value seed of crops like soybean, wheat, *etc*.

Madhya Pradesh became the first state in the country to have a private extension policy. Focus was on introduction of private extension services through building up private-public partnership in agricultural extension. The first Memorandum of Understanding (MoU) regarding implementation of public-private partnership in agriculture was signed by the Department of Agriculture with Dhanuka group for agricultural extension in Hoshangabad district of Madhya Pradesh. Another soybean-based model involving participatory approaches was that of ‚Gramin Vikas Trust' (an arm of Krishak Bharti Cooperative Ltd, KRIBHCO) jointly with the National Research Centre for Soybean (ICAR) which undertook knowledge and technology dissemination in tribal-dominated Jhabua district of Madhya Pradesh. The model was unique in the sense that it involved the farmers fully in assessing their needs and disseminating required knowledge and technology in agriculture. The village level para-extension workers came from among the villagers themselves and were called ―Jaankaars‖. They were, albeit, trained by the involved agencies. The Madhya Pradesh state also has built-in support of such para-extension workers, in their programmes, called as ―kisan bandhus‖.

The distinctive model of ‚Soy choupals' or ‚e-Choupals' mooted and implemented by Indian Tobacco Company (ITC) is well known as an early and successful use of ICTs in agriculture (Kumar, 2004; Rao, 2007; Tiwari 2008). e Choupals functioned for soybean crop in the state of Madhya Pradesh and, hence, were also called as ‚soy-choupals'. These e-Choupals were provided with internet connectivity with solar panel battery backup and VSAT (Very Small Aperture Terminal, a satellite communication system) equipment. Village internet kiosks were managed by farmers, called ‚sanchalaks'. This system (i) enabled the agricultural community's access to information on the weather and market prices, (ii) disseminated knowledge on scientific farm practices and risk management, and (iii) facilitated the sale of farm inputs and purchase of farm produce at the farmers' doorsteps. It served primarily as a direct marketing channel, reduced the transaction costs and eliminated many intermediaries (Adhiguru and Mruthyunjaya, 2004; Rao, 2007). e-Choupals eventually got extended to a range of crops and covered several states to become one of the largest initiative among internet-based interventions in rural India.

## Emerging new technologies in processing

Almost all soybean grain, barring farm-saved seed, reserves and stocks, is processed in solvent extraction plants to obtain soymeal and oil. Soymeal segment is the largest. Bulk of soymeal is consumed by poultry followed by the swine, beef, dairy, pet food and aquaculture industries. Protein derived from soybean is further processed to derive specialty products such as soy concentrate and soy isolates used for industrial, food and neutraceutical applications.

There are several emerging and new technologies. The food science related emerging technologies include pulsed electric field applications, high hydrostatic pressure technology, ultrasound and cold plasma that have already applications in the food industry and related sectors (Knorr *et al.,* 2011). High hydrostatic pressure technology for food preservation and quality retention are the most advanced. Pulsed electric field applications are on the verge of industrial use. Ultrasound has some non food safety applications and supercritical water and low temperature plasma treatment are in their developmental stage. Combinations of the above processes with either mild heat, with each other or with other means of food preservation agents (*e.g.* anti-microbials) are also being developed and tested. Currently, these technologies are mainly used and developed as alternatives to conventional thermal preservation methods such as pasteurization. However, they possess a great potential for food modification purposes and are generally more sustainable technologies than conventional thermal ones. Some recent attempts to re-introduce antimicrobial systems such as the use of ozone, UVC light, pulsed light or oxidized water still need to be validated regarding their feasibility and effectiveness.

For food uses, solvent extraction, particularly use of hexane, is being questioned on account of inflammatory nature of hexane solvent, its stated carcinogenicity and environmental hazard. Some of the current research and developments in soyfood technology are in the area of extrusion-expelling, super critical fluid extraction of oil, extruded soybased snack foods, expander technology, membrane technology, texturized soyproteins, coagulation and fermentation of soyproteins, chilling and freezing, electromagnetic waves, high pressure pasteurization and cooking, ohmic heating, retort and aseptic packaging among others. High shear extrusion does not require an external source of heat or steam because heat gets generated through friction in such extruder.

Fabricated foods and feeds from soybean and soybean derivatives using technologies like extrusion processing, modified atmosphere packaging and controlled atmospheric packaging are to be increasingly promoted as per changing consumer demand and ever increasing urbanization in the modern era. The change in the retail market has induced the change in soybean food industries. Specifically, the emergence and dominance of supermarkets in the retail market has motivated large soybean food makers to invest in new technologies to produce uniform and standardized products. Even traditionally soy-food using countries (for example Japan with traditional soyfoods namely, tofu, miso and natto) are changing the production and packaging to suit the product selling in big metropolitan stores, supermarkets and shopping malls or through large scale urban home delivery mechanisms and agencies. The product attributes needed and the emergence of new markets is affecting the structure of industry.

**The new era of biological/genetic modification, bioactive compounds and specialty soybean**

The genetically modified (GM) or transgenic soybean is already covering one billion hectares area globally (ISAAA, 2015). However, the biological world has gone ahead of genetically modified (GM) or transgenic crops that are developed using *Agrobacterium* or particle bombardment for crop improvement. A new crop biotechnology called ―genome or gene editing‖ is the recent development. There are different types of genome editing technologies. The most promising one is named CRISPR (Clustered Regularly Interspaced Short Palindromic Repeat). These new technologies consist of cutting the DNA at a pre-determined location and the precise insertion of the mutation, or single nucleotide changes at an optimal location in the genome for maximum expression. The real power of these new technologies is their ability to edit and modify single or multiple native plant genes (non-GM) coding for important traits. It is stated (ISAAA, 2015) that such developed improved crop varieties are not transgenics and retain their non GMO status. Products already under development using this technology include several major foods and feed crops like soybean (for improving oil quality). More complex traits, coded by multiple genes, like improved photosynthesis, are planned in near future. A trident confluence of the new breeding technologies (NBT) comprising transgenic development, genome editing, and use of microbes (plant microbiomes) as a new source of additional genes will be increasingly used to enhance productivity and quality of crop produce and products. In the long run, the genome editing technologies have the potential to settle the GMO *vs* non-GMO debate in Indian soybean scenario. Organic soybean producing large states like Madhya Pradesh in central India could have co-existence of organic and genome-edited improved soybean production in future for high yield and quality attributes.

Soybean derivatives contain many bioactive compounds that have health implications (El-Shemy, 2013). Several of them, when processed and consumed appropriately, could impart significant health benefits (Isanga and Zhang, 2008). Already much has been talked and published for and against the food uses of soybean (Fallon and Enig, 2000; Daniel, 2005; Adams, 2007; Robbins, 2012, and many others). Even the loudest anti-soy voices believe that traditional fermented soyfoods such as miso, natto, shoyu, and tempeh are good. Anti-nutritional factors can be taken care of by genetics and breeding, processing, bio-processing and food and feed modifications. Interest in food-grade soybeans and soy-derived neutraceuticals will grow and the needed technologies will evolve to address the encumbrances in the way of augmented food uses of soybean.

Agronomic characters, particularly yield and stress resistance, in soybean varieties are very important for their adoption by farmers, but lack or moderation of these variables, such as composition, functional characteristics (as in isoflavones) and absence of antinutritional compounds (as trypsin inhibitor) are now becoming desirable for their use by food industry. Such lines are available and some have been commercialized to an extent. However, these need to be further improved for yield and agronomic acceptance. There is less yet steady consumer demand for food-grade soybeans, including specialty varieties, organic, non-GMO, and Identity Preserved (IP) soybeans. Associated specific mechanisms have, therefore, to be put in place for segregation, identity preservation, a change in processing and packaging, certification and sustainable growing regimen.

As stated, many specialty soybean varieties for food uses have been developed globally. Indian scenario also has such soybean varieties. Lines namely, ‗NRC 101' and ‗NRC 102' which are free from ‗kunitz' trypsin inhibitor along with ‗IC 210', an indigenous line, which has high oleic acid (~42 %), and NRC 109 with null lipoxygenase-2 have been commercialized as well (Rani and Kumar, 2015; Kumar et al., 2010). Thus, the first laudable step towards promoting specialty soybean and carving a new commercial niche market has been taken in India. Production, use and export of organic soybean in/from central India are also on increase. Area to be brought under organic farming should be carefully identified rather than abruptly cannot be considered an exclusion factor, since other important claiming established high yielding regions. In order to meet the emerging demand for segregation and identity preservation, current infrastructure of solvent extraction plants will need to be modified or could co-exist with technologies like extrusion expelling.

The intellectual property rights (IPRs) have started covering wider areas than before leading to new IP-regime. The specialty soybeans in India are being produced under special agreements with private sector and thus indicating a need for different or hybrid intellectual property rights and mechanisms of benefit sharing. Contract farming for food uses could emerge as a potent tool to solve many problems such as those pertaining to farm-holding size, scale of enterprise, identity preservation, specialty soybean, organic farming, certification, technology adoption and overall farm and marketing management.

**Soybean oil in graphene production – An example of new soybean use**

A very recent report established an important use of soybean oil as a biomass precursor for the production of graphene, probably the hardest material with myriad potential uses. Graphene, an atomically thin film of crystalline carbon, is a highly promising nano-carbon material. Widely adopted techniques for the synthesis of such carbon nanostructures are primarily based on thermal chemical vapour deposition methods, in which purified gases are processed at very high temperatures (typically around 1,000 °C) over a prolonged period. The use of purified gases is, however, expensive, hazardous and requires extensive vacuum processing. A need was, therefore, felt to develop a process that is free of compressed gases for the production of graphene films. Recently, Dong Han Seo *et al.* (2017) of CSIRO, Australia, developed a method for single-step, rapid thermal synthesis of uniform and continuous graphene films in an ambient air environment, using a cheap and renewable form of biomass, soybean oil, as the precursor. Importantly, the method offers the scope to potentially address the critical roadblocks towards large-scale and efficient graphene manufacturing.

## Alternative protein sources of future for environmental and resource sustainability

Soybean is an established source of protein and currently dominates the plant protein market but it has to compete with other sources. Other than soymeal, there are several traditional plant protein sources of the world such as cottonseed meal, rapeseed meal, sunflower meal, groundnut meal, flaxseed meal, peas, faba beans and quinoa (*Chenopodium quinoa*). Besides plants, soybean has to compete with other non-plant protein sources. For example, soybean and fishmeal have traditionally been the main protein sources in poultry feeds. It would be desirable that vegetable protein from soybean and other plant sources is directly consumed by human beings as the conversion chain from vegetable protein to meat to humans is inefficient. Yet, world meat production is projected to double by 2050 particularly in developing countries like China and India where annual per capita consumption of meat has doubled since 1980. The proportion of land utilised to produce food for livestock to meet growing demand of meat is increasing. Future may see emergence of some yet uncommon alternative sources of protein *viz.*, cultured meat, mealworms and other insects, microalgae and cyanobacteria. Synthetic amino acids have a long history and with advancement in technology their incorporation may increase for value addition in food and feed.

There is a small but growing entomophagy and flexitarian sections of human communities using as many as 1900 species of insects as food worldwide. Many insects possess protein comparable with meat, essential amino acid levels comparable with soybean proteins along with sustainability and resource use efficiency features such as higher food conversion efficiencies, lower environmental impact and higher potential to be grown on waste streams (Ramos-Elorduy, 1997; Oonincx *et al.*, 2010; Lakemond *et al.*, 2013; van Huis, 2013). Mealworms, for example, have up to 48 per cent protein by weight and besides being taken as raw can be dried for grinding and added to the food. FAO has also been promoting insects for food and feed as viable protein source for humans as a way to reduce the environmental impact of meat.

Algae, both macroalgae and

microalgae categories, have potential food and feed uses. These have not only proteins but also vitamins and nutrients. Spirulina (a form of Cyanobacteria) and Chlorella (a single-cell green algae) are called as 'superfoods'. The future of microalgae is promising but presently these are too costly a protein source to effectively compete with other plant proteins.

Animal protein coming from livestock has encumbrances of being an inefficient food conversion chain and environmentally contributing about 18 per cent of global greenhouse gas (GHG) emissions, having 27% of the global water footprint and occupying 33 per cent of the global land use (Tuomisto and Teixeira, 2011). Cultured edible meat produced through *in vitro* tissue culture technologies, has a potential to overcome these concerns. Cultured meat production has a maximum potential reduction in GHG emissions and water footprint between 78-99 per cent and 89-96 per cent, respectively. If the same quantity of protein is replaced with soybean protein, even higher reductions in GHG emissions could be achieved, but the water footprint and land use would be higher compared to cultured meat. Cultured meat technology is futuristic and presently in a research stage. The first commercial product is predicted to be available within a decade. To start with, it will be desirable for mankind to move from consuming livestock protein to soyprotein directly for environment and resource sustainability.

**Present trends and perspective in soybean derivatives**

Soybean is the most globalized, traded and processed crop commodity. Soybean derivatives market can be divided (1) by type comprising (i) soybean, (ii) soymeal (soy milk and soy protein concentrate), and (iii) soy oil (soy lecithin), (2) by application comprising (i) feed, (ii) food, and (iii) other industries including biodiesel, soy-based wood adhesives, soy ink, soy crayons, soy based lubricants and many more, and (3) by lecithin processing comprising (i) water, (ii) acid, and (iii) enzyme, and (4) by region namely, North and South America, Europe, China, Japan, Southeast Asia and South Asia (India).

Wang *et al.* (2013) have concluded, using FAO data, that in the last 5 decades (1961-2010), the total production of soybean increased more than 8 times, mainly due to the increase of the cropping area, and contributed to the restructuring of agriculture in the world. The world soybean consumption patterns have changed over time; currently about 80 per cent went into processing industry and about 10 per cent is directly used for food purposes. The global export orientation ratio (EOR *i.e.*, total export in per cent of production) of soybean is higher than that of wheat, maize and rice. United States and South America have emerged as the main producers and exporters now while Asia particularly China has changed into the biggest importer. Soybean transportation profile shows that the main exporting ports are Mississippi Gulf Coast in US, Port of Rosario in Argentina, and Porto de

Santos and Porto de Paranaguá in Brazil. China is the largest importer of soybeans. China shares about 60 per cent of world imports almost all of which arrives in the port of Dalian. Modernization and up gradation of ports and industrialization of port-hinterlands for exports is, therefore, essential and is happening (*e.g.*, improvements in the Port of Rosario in Argentina, India's Sagar Mala Initiative, *etc.*).

Research and Markets (2015) estimates showed that soybean derivatives market was worth $ 176.92 billion in 2015, and projected to reach $ 254.91 billion by 2020, at a compound annual growth rate (CAGR) of 7.6 per cent. Asia-Pacific region was estimated to be the largest market and projected to grow to $ 125.85 billion by 2020, at a CAGR of 7.8 per cent. North America was estimated the second-largest market that is projected to grow at a CAGR of 9.2 per cent. USA, Argentina and Brazil are and will continue to be the top three producers and exporters of soybean and soymeal. Followed by these, Indian soy industry has also made a mark in national and global arena.

Leading companies of soybean derivatives market such as Bunge Ltd (U.S.), Archer Daniels Midland Company (U.S.), Louis Dreyfus Commodities (The Netherlands), Cargill, Incorporated (U.S.), Noble Group Ltd (Hong Kong), Wilmar International Limited (Singapore), Noble Group Ltd., CHS Inc., AG Processing Inc., Ruchi Soya Industries Limited (India), and Du Pont Nutrition and Health are some of the key players presently which employ various strategies for their growth and development. Expansion, diversification, joint ventures and acquisition have been the key moves undertaken by them. In Indian scenario some prominent soy processing industrial groups are Ruchi Soya Industries Limited (Ruchi Group of Companies), Prestige Feed Mills Ltd. (Prestige Group of Industries), Vippy Industries Ltd., Premier Industries Ltd., Divya Jyoti Industries, and several others. Most of them are based in Malwa (Madhya Pradesh) and also in other parts of the country. The soy-industry in totality manufactures high quality edible soybean oil, hydrogenated 'vanaspati', bakery fats and soyfoods (soy-chunks, -granules, and –flour) and export soymeal, lecithin and other derivatives. In the year 2010, the industry was found offering employment to 12 lakhs people and estimated to go up to 15 lakhs in 2015 and up to 18 lakhs in the year 2020 although the trickle down is much more than this.

In India, there are about 120 soybean processing plants. The installed capacity of over 25 million tonnes is much more than the present normal annual soybean production of 12 to 14 million tonnes. India normally exported soymeal plus other soy-products valued at around US $ 2.5 billion annually till 2013-14. In the year 2012-13, soymeal alone was exported to the tune of about 3.439 million tonnes with a value of Rs. 10,050 crores. Soymeal export for 2011-12 and 2010-11 was 3.829 and 3.838 million tonnes respectively (Source: SEA). After 2013-14, the exports declined severely due to decrease in soybean production,

stiff global competition and several other reasons. Domestic consumption as feed (~ 3.9 million tonnes annually) is also substantial. Soyoil (~1.7 million tonnes per year) is almost entirely consumed domestically, with additional soyoil being imported for meeting the domestic demand. Owing to this very reason, the use of soyoil for biodiesel is unlikely in India. Consequently, the demand for corn ethanol could be the reason for competition between soybean and corn for acreage in US and other countries but not so in India.

**Import of edible oil**

The import of edible oil in India has soared in order to meet the felt need of a growing population with rising income levels. In the year 2014-15 (November-October), India imported about 14.4 million tonnes of edible oil (crude plus refined), with a huge import reliance of more than 60 per cent of our requirement (about 22 million tonnes) and with a high value of about US $ 10 billion. Palm oil constitutes more than 60 per cent of the import, mainly from Indonesia and Malaysia, followed by soybean oil mainly from Argentina and Brazil and sunflower oil mainly from Ukraine and Mexico. Soybean has come as an additionality to Indian oilseed scenario and in spite of it being more a protein than oilseed crop, it contributes about 1.7 million tonnes of edible oil every year. Efforts are being made on several fronts to increase soybean production. Regulatory aspects and tariff structures, however, have become sensitive for domestic companies. Organisations such as the Solvent Extractors' Association of India (SEA), the Central Organisation for Oil Industry and Trade (COOIT) and the Soybean Processors Association of India (SOPA) along with others have made suggestions from time to time in this regard. A high duty on import, particularly on a single large quantity, may probably be imposed. Soybean oil has a 45 per cent bound duty tariff under WTO. Current applied duties on crude palm oil have been decreased from 12.5 per cent to 7.5 per cent and on refined palm oil from 20 per cent to 15 per cent in the year 2016. A desirable tariff difference between crude and refined oil (a duty differential desirably higher than the present 7.5 %), particularly in case of palm oil, should be there so that domestic oil refineries, using crude imported oil, survive and thrive well.

A view may be taken as to whether we may offset the oil import bill by the export-earnings of oilseed sector itself, as was so decades ago. This is different from being self-sufficient *sensu stricto*. Globalisation of food commodities is taking place at a large scale, disconnecting production and consumption. Some deficient countries may use the land abroad to virtually increase their agricultural land and production. This is referred to as ‗virtual land use' or ‗displaced land use' (Rulli *et al.*, 2013; Yu *et al.*, 2013) and is one of the possible ways-out besides increasing the system-wide output.

**Industry slowdowns due to uneven and**

**low supply of raw material – instability in Indian soybean production**

Soybean production fluctuates and experiences serious dips, mostly due to weather aberrations. Due to this, soybean farmers and industry have suffered most several times but two hard periods of slowdown that cannot be forgotten are (i) for 3 years from 2000-01 to 2002-03 and (ii) for 3 years from 2013-14 to 2015-16. Instability in soybean yields was found to be high (Sharma, 2016) in all major soybean growing states (ranging from 21 to 43 %) as well as on all India basis (about 20 %). Soybean is largely a commercial crop and an agro industrial venture. Unstable and low supply of raw material to industry leads to depression in export of soymeal and other products as also in their availability for domestic consumption. It, then, becomes hard to compete globally on price in spite of non-GMO status and other USPs (unique selling points/propositions) of Indian soymeal. Resilience shown by soybean industry and re-gaining its strength (now in 2016-17) is praiseworthy. Domestic sale of oil also compensates to an extent for such volatility in soymeal prices but this recourse needs policy support as stated above. Availability of soybean as raw material for industry, adherence to quality and meeting global standards will decide the viability of our export slots and their size.

Climate change has been held responsible for instability in soybean production. The studies, however, show that in the backdrop of climate change, soybean fairs a degree better than other major crops. Rising carbon-dioxide may have some positive effect on soybean yield although a range of yield effects has been observed (Adams *et al*. 1998). Collateral improvement in the photosynthesis/transpiration ratio, amounting to water-use efficiency, would offset some of the negative effects of global warming in a C3 plant species like soybean. Keeping aside this argument, soybean research and development organizations should develop climate resilient high yielding varieties and commensurate technologies and take them to farm level to provide stability and increase in production. Probably one million hectares more can be added in near future to the present Indian soybean area of about 12 million hectares but production intensification rather than area increase will be the principal means of meeting future demand. A yield gap of about one tonne per hectare already exists to be filled up by technology adoption. It is not deft to compare Indian soybean yield level with that of US and other countries owing mainly to major differences in the length of crop duration. Indian soybean, particularly in the major central Indian region, is grown under a very short duration of about 95 days after which other crops follow, the crop sequence being soybean-wheat or even intensive soybean-potato-late sown wheat with high system efficiency rather than high yield of a single crop. All comparisons should be made on the basis of per day productivity or overall system efficiency. The differences will, then, be not so wide. Also, there are several studies to show that the yield gap cannot be fully bridged in a rainfed crop like soybean. Still, farm-level yields need to be raised. Based on available yield gap (one tonne per hectare) and likely extent of bridging it up (60 to 80%), it has been estimated that the national average yields can be increased feasibly to reach about 1.6 tonnes per hectare and then with some more efforts to 1.8 tonnes per hectare despite the short growing period available in central India (Tiwari, 2001;

Tiwari, 2014). The National Mission on Oilseeds and Oil Palm (NMOOP) since 2014 (a new form of the legendary Technology Mission on Oilseeds that was established in May 1986), agricultural universities, ICAR centres and several other public and private initiatives/organisations are striving for increasing the production of oilseeds and other edible oil sources.

Future will see increased use of automated systems, precision agriculture, use of nanotechnology, GPS, remote and satellite-monitoring, market information systems and use of sensors and drones in farm operations. Meeting the needed scale of economy can be achieved by contract farming and by establishing famer producer companies and farmers' seed cooperatives. Input cost moderation, facilitated availability of inputs, crop insurance, easy loans, support price and market-intervention are being undertaken and should ideally contribute towards measures taken for improving terms of trade in farming.

Keeping in view the global export of Indian soy-derivatives, public sector guidelines such as Codex Alimentarius will remain important to ensure food safety, while private sector standards such as Global Good Agricultural Practices (GAP) may become increasingly significant. Expanding Asian markets may embrace these guidelines.

## Sustainably produced food with fair trade - A new Consumer preference

Worldwide, an increasing number of consumers are showing a greater interest in the holistic quality attributes of the food that they consume. Beyond the immediate issues of food safety, a competitive price and the experiential quality attributes, the consumers are becoming more concerned about the manner in which their food has been produced. Besides nutritional and sensory aspects, urban consumers are increasingly inclined towards fair trade, sustainable, non-GMO and organic certified products when choosing foods and evaluating food quality. Certified non-GMO soybean meal and other soy products (certified by CERT ID for example) are already in demand (Freire A, 2013). Present major international voluntary sustainability standards include the Danube Soya Initiative, Fairtrade, the Round Table on Responsible Soy (RTRS), ProTerra, Organic, the International Sustainability and Carbon Certification (ISCC) and the Round Table on Sustainable Biomaterials (RSB). Brazil, China and Argentina are the largest standard-compliant producers (SSI Review, 2014). The multi-stakeholder association namely, Fairtrade Labelling Organizations International had split, in the year 2004, into Fairtrade International and FLOCert. Buyers of sustainably produced and fair-trade products state that it makes them sure that the environmental care has been taken in production along with no residual pesticides, no use of child labour, *etc*. High-priced products with such specific tags on sustainably produced and fair trade products are now in vogue in developed countries. In developing countries like India also, sustainability certification programmes are being implemented. For example, South and South East Asia network of Solidaridad (a development agency with its headquarters in Utrecht, Netherlands) has a Farmer Support Program with its seven NGO partners in operation since 2013 in three major soybean growing states of India namely, Madhya Pradesh, Maharashtra and Rajasthan.

Despite a large area under genetically modified (GM) soybean in USA, Brazil and Argentina, future demand for non-GM soybean will continue in some regions like EU. This could be advantageous for India where no GM soybean is yet grown and some area is even under organic soybean. Also, value of the produce/products and income generated could decide the farming options for a crop and the variety within a crop in future rather than yield alone. These recent consumer preferences for fair trade and sustainable production point out that we need to expand single-minded yield-centric approach to address both sustainability and farm prosperity concerns in a system-wide mode.

**Making economic activities upsurge –**

**Incentivisation and Trickle down**

The trickle down for welfare of the society at large, as happened in case of soybean revolution (Badal *et al.*, 2000), will continue if the companies and stakeholders are incentivised to generate income, raise output and create better and more opportunities for entrepreneurship, self-employment and also jobs. Optimisation of tax-revenue and maximisation of economic activities are needed for continued growth and national prosperity. It is a known fact that economic activities upsurge and people (farmers, industrialists and entrepreneurs) work hard and efficiently when terms of trade are favourable and input-cost and taxation are in the desirable range. That range or point may or may not be consistent with that of the Laffer Curve (Fullerton, 2008; Laffer –The Heritage Foundation, 2016; Trabandt and Uhlig, 2011), which is otherwise debatable. Above all, it is not the tax rate alone that matters. In nutshell, what really matters for doing business and making investment is the potential profit (an incentive in itself) along with the degree of uncertainty in earning that profit. Some recent and upcoming reforms and policy decisions in India include the Goods and Services Tax (GST) legislation, Land Acquisition legislation, new Foreign Trade Policy adopted in 2015, trade facilitation measures (www.indiatradeportal.in), integration of 20 services for obtaining government clearances (through eBiz single window portal), timely tax and tariff rate reviews and other initiatives of the Ministry of Commerce and Industry and related ministries, which are significant and noteworthy. Continuation of a conducive policy environment and dynamic trade facilitation are needed to bring upsurge in the economic activities including those concerning soy production and soy-derivatives business.

The pace of development has increased tremendously of late and the world is changing fast. New production and processing technologies, new and customized products, changing global standards and demands, operational innovations for slashing logistics and other costs, diversification, convergence and co-existence of interests and companies, and other efficient ways of doing business like increased use of IT based applications and cyber-physical systems for execution of complete global value chain will continue to have widespread and expanding effect on production and commerce of soybean. The trends and minutiae pointed out in this article show that soybean, being a major global industrial crop, stands to benefit from the ramifications of the new industrial era. Present trends, volatilities, slowdowns, challenges faced and associated desiderata have been spelt out accordingly.

## ACKNOWLEDGEMENT

The author is grateful to experts from several organisations especially IISR (ICAR), SOPA and some group of industries in Indore for their direct and indirect support. He is also thankful to the two anonymous referees for their suggestions.

15